\documentclass[epj]{svjour}
\usepackage{graphics}
\begin{document}
\title{Scaling laws and higher-order effects 
in Coulomb excitation of neutron halo nuclei}
\author{S. Typel\inst{1,2} \and G. Baur\inst{3,4}% etc
}                     % Do not remove
\offprints{S.Typel}          % Insert a name or remove this line
\institute{Excellence Cluster Universe, Technische Universit\"{a}t M\"{u}nchen,
Boltzmannstra\ss{}e 2, D-85748 Garching, Germany
\and
Gesellschaft f\"{u}r Schwerionenforschung (GSI) mbH, Planckstra\ss{}e
1, D-64291 Darmstadt, Germany
\and
Institut f\"{u}r Kernphysik, Forschungszentrum J\"{u}lich,
D-52425 J\"{u}lich, Germany
\and
J\"ulich Centre for Hadron Physics, Forschungszentrum J\"ulich, 
D-52425 J\"ulich, Germany} 
\date{Received: date / Revised version: date}
\abstract{
%Insert your abstract here.
Essential properties of
halo nuclei can be described in terms of a few 
low-energy constants. For neutron halo nuclei, analytical results can be 
found for wave functions and electromagnetic transition
matrix-elements in simple but well-adapted models. 
These wave functions can be used to study nuclear reactions; 
an especially simple and instructive example is Coulomb excitation. 
A systematic expansion in terms of 
small parameters can be given.
We present scaling laws for excitation amplitudes and cross sections.
The results can be used to analyze experiments
like ${}^{11}$Be Coulomb excitation. They also serve as benchmark tests
for more involved reaction theories.
\PACS{
      {25.70.De}{Coulomb excitation}   \and
      {23.20.Js}{Multipole matrix elements} \and
      {27.20.+n}{$6 \le A \le 19$}
     } % end of PACS codes
} %end of abstract
\maketitle
\section{Introduction}
\label{intro}
%Your text comes here. Separate text sections with
%\section{Section title}
%\label{sec:1}
%and \cite{RefJ}
%\subsection{Subsection title}
%\label{sec:2}
%as required. Don't forget to give each section
%and subsection a unique label (see Sect.~\ref{sec:1}).

Exotic nuclei are available as secondary beams at many radioactive beam
facilities around the world. These unstable nuclei are generally 
weakly bound with few, if any, excited states. 
A well developed method to study halo nuclei
is Coulomb excitation. An instructive example is the excitation
of the $1/2^{-}$ bound state in ${}^{11}$Be from the $1/2^{+}$ ground
state \cite{xcdcc,ganil,riken,fau}.

Halo nuclei are a low-energy phenomenon and can be described effectively
in terms of a few low-energy parameters. The ratios
of core size to the sizes of the halo states
(to be defined in eq. (\ref{expapara}) below) serve
as small expansion parameters. Wave functions and matrix elements can be given 
in terms of these parameters.
These wave functions can be used in reaction
models and simple and realistic formulae are obtained.

In a recent paper the $B(E1;1/2^{+} \rightarrow 1/2^{-})$ strength
for ${}^{11}$Be has been determined from intermediate energy Coulomb excitation
measurements \cite{xcdcc}. 
In order to analyse such kind of data in terms of electromagnetic 
matrix elements
it is certainly necessary to use rather sophisticated codes which 
take higher-order electromagnetic and nuclear effects into account.
These codes can be checked and validated by
comparing their results to limiting cases where analytical 
results can be obtained, e.g., pure
Coulomb excitation that shows some rather simple features.
This can provide a useful guide for more sophisticated 
approaches. It is the purpose of this paper to 
provide such analytical results.
In the theoretical analysis of \cite{xcdcc}, 
quite a complicated XCDCC method was used,
which is required for a quantitative description of the experimental
data, but it may tend to obscure the understanding of the
simple physical mechanism of Coulomb excitation.
${}^{11}$Be is an archetype of a halo nucleus with a ${}^{10}$Be core and 
a single halo neutron in the $2s_{1/2}$ state. There is a strong $E1$
transition
to the $1/2^{-}$ bound state, itself a $p$-wave neutron halo state. This dipole
transition was previously studied by Coulomb excitation at GANIL, RIKEN, and 
MSU \cite{ganil,riken,fau}.  

There are two somewhat separated questions:
the influence of nuclear excitation in grazing collisions and 
higher-order 
electromagnetic effects in distant collisions where nuclear interaction
effects can safely be neglected. We deal here with the second question: 
higher-order electromagnetic excitation. This problem was studied 
for the $^{11}$Be case in the 
mid-nineties by two groups \cite{tyba95,bch}. 
%{\tt (Hier k\"{o}nnte
%  man f\"{u}r den allgemeinen Fall eigentlich noch mehr angeben.)}
In view of recent advances in the description of the electromagnetic properties
of halo nuclei, see e.g. \cite{tyba04,tyba05,nlv,Liu03}, 
it seems appropriate to update
this work using the recent analytical results for halo wave functions
and present scaling laws for Coulomb excitation of halo nuclei.
Effective field theory methods are also applied successfully to halo nuclei,
see \cite{Bed03,Ber02}.

In  section 2  we give the main theoretical formulae for 
our model of the Coulomb excitation of neutron halo nuclei
from $s$- to  $p$-states.
In section 3 scaling rules are discussed. Then we  
give an application to the case of ${}^{11}$Be.
This can serve as a benchmark for 
more involved studies like XCDCC or time-dependent approaches \cite{tidep}. 
%({\tt Refs. D. Baye et al. \dots}). 
Conclusions are given in section 4.
A preliminary account of part of this work was published in the 
proceedings of the conference 'Nuclear Physics 
in Astrophysics III' in Dresden, March 2007 \cite{baty08}. 

\section{Analytically solvable model for 
Coulomb excitation of neutron halo nuclei}

%Halo nuclei are low-energy phenomena and thus
A few low-energy parameters are sufficient to characterize
halo nuclei. Let us consider the single-particle excitation of
a neutron from a ground state $i=0$ to a bound excited state
$i=1$ with neutron separation energies $E_{i}>0$. 
(For ${}^{11}$Be we have $E_{0}= 504$~keV and $E_{1}=184$~keV.)
The size of the
single-particle wave functions is determined by the
bound-state constants
$q_{i} = \sqrt{2\mu E_{i}}/\hbar$ with the reduced mass
$\mu=m_{n}m_{c}/(m_{n}+m_{c})$ of the nucleon+core system.
With the radius $R$ of the core beyond which the nuclear interaction is 
assumed to vanish, it is possible to introduce
dimensionless parameters
\begin{equation}  \label{expapara}
\gamma_{i} = q_{i} R
\end{equation}
that are a measure
for the ratio of the core size to the size of the neutron wave functions.
These parameters are small for halo nuclei, see, e.g.,
\cite{tyba04,tyba05}, and can be used as convenient expansion
parameters. In this spirit, we use this single-particle model 
for the $j_{i}^{\pi} = 1/2^{+}$ and $j_{f}^{\pi} = 1/2^{-}$ 
states in order to evaluate
the matrix element for the $E1$ electromagnetic excitation. It is
dominated by the exterior contributions.

\subsection{Halo wave functions and matrix elements}

The single-particle wave functions are given by
\begin{equation}  \label{full}
 \Phi_{i}(\vec{r}) = \frac{f_{i}(r)}{r} 
\mathcal{Y}_{j_{i}m_{i}}^{l_{i}s}(\hat{r})
\end{equation} 
for the ground state ($i = 0$) and the excited state ($i=1$). 
The angular dependence is described by the 
spinor spherical harmonics $\mathcal{Y}_{jm}^{ls}$ ($s=1/2$). 
We use the results
and notation of \cite{tyba05} assuming zero spin of the core.
The radial wave functions
of the two states are given in the exterior
region ($r>R$) by  
\begin{equation}  \label{ext0}
f_{0}(r) = C_{0} q_{0}r  h_{0}^{(1)}( i q_{0} r)
 = - C_{0} \exp(-q_{0}r)
\end{equation}
and 
\begin{eqnarray} \label{ext1}
f_{1}(r) & = & C_{1} i q_{1}r  h_{1}^{(1)}( i q_{1} r)
 \\ \nonumber & = & 
  -C_{1} \exp(-q_{1}r) \left( 1 + \frac{1}{q_{1}r} \right)
\end{eqnarray}
with Hankel functions $h_{l}^{(1)}$ of imaginary argument.
Both states are halo states and the normalization 
constants are given in the halo limit by $C_{0}= \sqrt{2q_{0}}$ and $C_{1}
=\sqrt{2q_{1}^{2} R/3}$, respectively
\cite{tyba05}. 

The parameters $q_{0}$ and $q_{1}$ of the shallow bound
states in a halo nucleus are closely
related to the scattering length $a_{l}$ and the effective range
parameter $r_{l}$ for partial waves $l \ge 0$ in the effective
range expansion $k^{2l+1} \cot (\delta_{l}) = -1/a_{l} + r_{l}k^{2}/2
+ \dots$. The S matrix
$S_{l} = [\cot (\delta_{l}) + i]/[\cot (\delta_{l}) - i]$ has a pole
at $k = ik_{B}$, i.e.\ $\cot [\delta_{l}(k_{B})] = i$ for a bound
state with $k_{B} = q_{i}$. This gives the desired
relation
\begin{equation} \label{efr}
 (-1)^{l+1} q_{l}^{2l+1} = - \frac{1}{a_{l}} - \frac{1}{2}
 r_{l}q_{l}^{2} + \dots \: .
\end{equation}
Now there is a difference between $l=0$ and $l > 0$. For $l=0$ and
small $q_{0}$ we have in lowest order the relation $q_{0} = 1/a_{0}$
and the effective-range term is a small correction. For $l>0$ the
left hand side of (\ref{efr}) is smaller than the first two terms 
individually on
the right hand side and we have an enhanced value (as compared to
``dimensional considerations'') of the scattering length that is given
in lowest order by $a_{l} = -2/(r_{l}q_{l}^{2})$ (``fine tuned'').
In \cite{tyba04} the scattering length in the $l=1$, $j=1/2$ channel
of ${}^{11}$Be was determined to be $457(67,-66)$~fm${}^{3}$. With
$q_{1} = 0.0895$~fm${}^{-1}$ in this channel we find an
effective-range parameter $r_{1} = -0.547$~fm$^{-1}$ of ``natural order''
$R^{-1}$.
 
With the radial wave functions (\ref{ext0}) and (\ref{ext1}) we  calculate 
the $B(E1)$ value for the $1/2^{+} \rightarrow 1/2^{-}$-transition
as well as the 
higher-order effects in electromagnetic excitation. 
We propose this to be a model study and leave the spectroscopic factors
equal to one. (They could be adjusted, which would result in a 
quasi-realistic description of the ${}^{11}$Be system for our
purpose.) 
The $B(E1)$ value is given by
\begin{equation}
 B(E1)= \frac{1}{4\pi} \left(Z_{\rm eff}^{(1)} e\right)^{2}
 \left|R_{01}^{(1)}\right|^{2} \: ,
\end{equation} 
where $Z_{\rm eff}^{(1)}=Z_{c} m_n/(m_n + m_c)$ is the
dipole effective charge number and the radial dipole integral
is given by
\begin{eqnarray} \label{radi}
 R_{01}^{(1)} & = & \int_{0}^{\infty} dr \: f_{1}^{\ast}(r) r f_{0}(r)  
 \\ \nonumber & = & 2 \sqrt{\frac{\gamma_{0}}{3}} 
 \frac{(\gamma_{0} + 2\gamma_{1})}{(\gamma_{0} + \gamma_{1})^2} R
 \: .
\end{eqnarray}
In the present paper we are only interested in the 
halo limit, i.e., we keep only the lowest-order term in
the expansion in $\gamma_{0}$ and  $\gamma_{1}$. 
In this approximation we can use the exterior radial wave functions,
eqs. (\ref{ext0}) and (\ref{ext1}) in the 
integral eq.\ (\ref{radi}) down to $r=0$.
The correction terms from using the correct interior
wave functions are of higher order in the expansion parameters
$\gamma_{0}$ and $\gamma_{1}$.
For $R \rightarrow 0$ the radial dipole integral goes to zero because the 
normalization of the $p$-wave function tends to zero in this limit.
Thus $R$ must be kept finite;
we choose $R= 2.78$~fm as a realistic value for ${}^{11}$Be \cite{tyba04}.
This value determines the asymptotic normalization of the 
$p$-wave bound state. We find  $B(E1)=0.193$~e$^{2}$fm$^{2}$,
to be compared to the  value of $B(E1)=0.105(12)$~e$^{2}$fm$^{2}$ obtained from
an analysis of the GANIL data, see
\cite{xcdcc}. This value is consistent with other Coulomb dissociation 
experiments at RIKEN and MSU
and the value obtained by the Doppler shift attenuation method \cite{mill}.

\subsection{Coulomb excitation of neutron halo nuclei and scaling
  laws}

We treat electromagnetic excitation in the semiclassical approximation.
For high beam energies the classical trajectory can be taken as a 
straight line with impact parameter $b$.
In the sudden approximation one can take higher-order effects into
account in a convenient way. Like the related Glauber approximation
it is applicable for high beam energies and low
excitation energies. This is reasonably well fulfilled for ${}^{11}$Be,
even at GANIL energies.  
In \cite{tyba95} 
higher-order effects in the electromagnetic excitation of ${}^{11}$Be  
to the $1/2^{-}$ bound state were studied using the sudden approximation.
In \cite{bch} a continuum discretized coupled channels approach was adopted. 
In \cite{tyba01} an analytically solvable model for higher-order 
effects in the electromagnetic dissociation 
of neutron halo nuclei was presented.
In that work, there was only the transition from an $s$-wave bound
state to the continuum.
Now we consider the case where there is, in addition, 
a $p$-wave bound state, as it is the case in ${}^{11}$Be.
In the sudden approximation the dipole excitation amplitude
from the ground state $i$ to the final state $f$ is  given by 
\begin{equation} \label{asudden}
a_{\rm sudden}= \langle f |\exp\left(- i \vec q_{\rm Coul} \cdot \vec{r}
\right) |i \rangle
\end{equation}
where the transfered momentum is
\begin{equation}
 \hbar \vec{q}_{\rm Coul} = \frac{2 Z Z_{\rm eff}^{(1)}e^2}{v b}
 \vec{e}_{z} \: .
\end{equation}
The impact parameter $b$ is chosen to be in the $z$-direction;
this is convenient for the following calculation using polar coordinates.
The target charge number is denoted by $Z$ and $v$ is the beam velocity. 
The dipole approximation is quite
well fulfilled, since the dipole effective charge $Z_{\rm eff}^{(1)}$
%=\frac{Z_{c} m_n}{m_n + m_c}$ 
is much larger than the 
corresponding quadrupole charge. The neutron and core masses
are denoted by $m_{n}$ and $m_{c}$ respectively, the charge of the core
is given by $Z_{c}$.
The sudden approximation
is applicable for $\xi \equiv \omega b/v \ll 1$ , where  
$\hbar\omega = E_{0}-E_{1}$ 
is the excitation energy. Even for the comparatively
low GANIL energies of about 40~MeV/nucleon
this is reasonably well fulfilled.
The most important intermediate states are expected to be
in the low-energy continuum, where the dipole strength has a peak,
at around 1~MeV excitation energy. Thus the adiabaticity condition
$\xi \ll 1$ will also apply to the important intermediate states.
We note that for the virtual excitation of the high-lying 
giant dipole state the opposite limit $\xi \gg 1$ is realized,
this would lead to a real polarization potential, see e.g. \cite{alwi}.
It influences the classical trajectory of the projectile
and it is not important in our context.
It is also neglected in our model space.
The sudden approximation has the advantage 
that intermediate states are treated by closure, thus one only needs a 
model for the initial and final states, and not for all the 
intermediate states. 
In lowest order in $q_{\rm Coul}$ the first-order dipole approximation is 
obtained corresponding to a single-photon exchange between target and
projectile.
It is shown in \cite{tyba95} that third-order
$E1$ excitation is more important than second-order $E1$-$E2$ excitation.
The role of higher multipole excitations in the electromagnetic excitation
of one-neutron halo nuclei is also studied in \cite{cfv}.

In the appendix we show that
\begin{equation} \label{exact}
a_{\rm sudden} = (-1)^{j_{i}+m_{i}}
  \frac{C_0 C_1}{2i} \left[ \frac{1}{q_{\rm Coul}} I_{1}(z) +
\frac{1}{q_{1}} I_{2}(z)\right]
\end{equation}
where 
\begin{equation}
 I_{1}(z) = 2\left[1-\frac{\arctan(z)}{z} \right]
\end{equation}
and
\begin{equation}
 I_{2}(z) = \left(1+ \frac{1}{z^2}\right) \arctan(z) - \frac{1}{z}
\end{equation}
depending on the dimensionless parameter
\begin{equation} \label{defz}
 z = \frac{q_{\rm Coul}}{q_{0}+q_{1}}
 =\frac{q_{\rm Coul} R}{\gamma_0 + \gamma_1} \: .
\end{equation}
This quantity is proportional to the usual strength parameter
\begin{equation} \label{chidef}
\chi^{(1)}= \frac{Ze \langle f || \mathcal{M}(E1)|| 
 i \rangle}{\hbar v b}
\end{equation}
that characterizes the importance of higher-order effects in Coulomb
excitation processes. We note that this parameter 
becomes large in the halo limit ($\gamma \rightarrow 0$).
In contrast to $\chi^{(1)}$, the adiabaticity parameter 
$\xi= \hbar \omega/(vb)$
with the excitation energy $\hbar \omega = \hbar^{2}
(q_{0}^{2}-q_{1}^{2})/(2\mu)$ becomes small in the halo limit and
the application of the sudden approximation (\ref{asudden}) is very well
justified.

We can expand the excitation amplitude
 (\ref{exact}) in powers of $z$
with the help of the power series of the  $\arctan$ function.
This power series expansion converges for $|z| \le 1$  and one obtains 
\begin{equation} \label{asuddenseries}
 a_{\rm sudden} = a_{1} + a_{3} + \dots
\end{equation}
with the lowest-order term
\begin{equation}
 a_{1} =  (-1)^{j_{i}+m_{i}} \frac{2}{3i} \sqrt{\frac{\gamma_{0}}{3}}  
 \frac{\gamma_{0}+2\gamma_{1}}{\gamma_{0}+\gamma_{1}} z
\end{equation}
and the next-to-leading order term
\begin{equation}
 a_{3} =  (-1)^{j_{i}+m_{i}} \frac{2i}{15} \sqrt{\frac{\gamma_{0}}{3}}  
 \frac{\gamma_{0}+4\gamma_{1}}{\gamma_{0}+\gamma_{1}} z^{3} \: .
\end{equation}
The even terms vanish due to the parity selection rule.
Note that in the sudden approximation (\ref{asudden}) there is no change
in the magnetic quantum number, since the vector $\vec{q}_{\rm Coul}$
is in the $z$-direction.

\begin{figure}
%\resizebox{0.75\textwidth}{!}{%
\resizebox{0.48\textwidth}{!}{%
\includegraphics{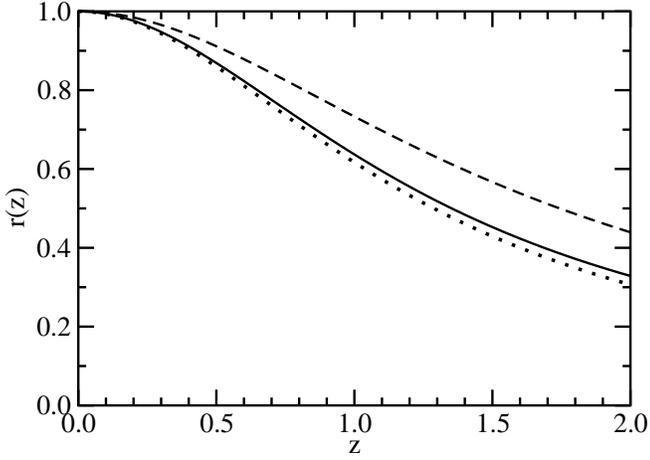}
}
\caption{Reduction factor $r(z)=\left|a_{\rm
      sudden}\right|^{2}/\left|a_{1}\right|^{2}$
as a function of the scaling parameter $z$ as defined in eq.\
(\ref{defz}) for values of $0$ (dashed line), $0.604$ (solid line)
and $1$ (dotted line) of the ratio 
$\gamma_{1}/\gamma_{0} = \sqrt{E_{1}/E_{0}}$.
}
\label{fig:1}       % Give a unique label
\end{figure}

Following Ref.\ \cite{tyba95} we can introduce a reduction factor 
\begin{equation}
  r (z) = \frac{\left|a_{\rm sudden}\right|^{2}}{\left|a_{1}\right|^2} 
 \approx 1 - \frac{2}{5} 
 \frac{\gamma_{0}+4\gamma_{1}}{\gamma_{0}+2\gamma_{1}} z^{2} + \dots \: .
\end{equation}
In figure \ref{fig:1} the dependence of the reduction factor $r$ on $z$
is shown for the present case with 
$\gamma_{1}/\gamma_{0}=\sqrt{E_{1}/E_{0}}\approx 0.604$ by the solid
line. We find $r(z) \approx 1-0.619~z^{2}$ in lowest order in $z$.
It seems of interest to compare the results of \cite{tyba95} 
with the present one.
In \cite{tyba95} two different models (denoted by $I$ and $II$) were
used. Expanding in the strength
parameter $z$ we get
%, using  the present notation,
$r^{I}(z) =1-6z^{2}/5 + \dots $, and 
$r^{II}(z) = 1-12z^{2}/5 + \dots$. 
The present higher-order corrections are smaller than the one 
found in \cite{tyba95}. In that reference rather crude models were used for
the first excited $1/2^{-}$-state; 
those wave functions extended much further out than
the present wave function (\ref{ext1}), which takes the angular momentum 
barrier into account properly. In the higher-order terms the 
importance of the outer region is enhanced due to the weighting
with a higher power of r. It should be recalled that the main aim of 
\cite{tyba95} was to provide an upper limit for higher-order effects in
order to understand the results of \cite{ganil}. In the meantime,
%similar 
other Coulomb excitation experiments were performed \cite{xcdcc,riken,fau}
that clarified the situation.

In the general case of a Coulomb excitation from a bound $s$-wave to a
bound $p$-wave neutron halo state, the energy $E_{1}$ of the excited
state is limited from below by zero (state at the breakup threshold) 
and from above by $E_{0}$ (zero-energy excitation). 
This corresponds to the intervall
$[0,1]$ for the ratio $\gamma_{1}/\gamma_{0}$. The two limiting
cases are depicted in figure \ref{fig:1} by dashed and dotted lines.

The excitation probability is given by 
\begin{eqnarray}
 P(b) & = & \frac{1}{2j_{i}+1} \sum_{m_{i}m_{f}} 
 \delta_{m_{i}m_{f}} \left| a_{\rm sudden} \right|^{2}
 \\ \nonumber & = & 
  P_{\rm LO}(b) + P_{\rm NLO}(b) + \dots \: .
\end{eqnarray}
The lowest-order term 
\begin{equation}
P_{\rm LO}(b) =  \frac{4 \gamma_{0} (\gamma_{0}
 + 2\gamma_{1})^{2}}{27 (\gamma_{0}+\gamma_{1})^{2}}
 z^{2}
 \equiv \frac{A_{2}}{b^{2}}
\end{equation}
in the excitation probability is proportional to $z^{2}$.
The most important higher-order contribution comes from $a_{3}$.
Its interference with the lowest-order term $a_{1}$ leads to the next term
\begin{equation}
P_{\rm NLO}(b) = -\frac{8\gamma_{0}(\gamma_{0} +2 \gamma_{1})
(\gamma_{0} + 4 \gamma_{1})}{135(\gamma_{0} + \gamma_{1})^{2}} z^{4}
\equiv - \frac{A_{4}}{b^{4}} 
\end{equation}
in the expansion of $P(b)$ in $z$, of the order of $z^{4}$. 
The constants $A_{2}$ and $A_{4}$ can be written as 
\begin{equation}
 A_{2} = \frac{4 \gamma_{0} (\gamma_{0} + 2\gamma_{1})^{2}}{27 
 (\gamma_{0} +\gamma_{1})^4} 
 \left(\frac{2\eta m_{n}}{m_{c} + m_{n}}\right)^{2} R^2
\end{equation}
and 
\begin{equation}
 A_{4} = \frac{8\gamma_{0}(\gamma_{0} +2 \gamma_{1})
(\gamma_{0} + 4 \gamma_{1})}{135(\gamma_{0} + \gamma_{1})^{6}} 
 \left(\frac{2\eta m_{n}}{m_{c} + m_{n}}\right)^4  R^4
\end{equation}
where the Coulomb parameter is given by $\eta=Z Z_{c} e^{2}/(\hbar v)$.
Total cross sections are obtained by integration over the impact 
parameter, starting from a minimum impact parameter $b_{\rm min}$.
The sudden approximation fails for large impact parameters, and 
an adiabatic cut-off
$b_{\rm max}=\gamma_{\rm beam} v/\omega$ 
with the Lorentz-factor
by $\gamma_{\rm beam}=1/\sqrt{1-(v/c)^2}$
has to be introduced for the lowest-order result. 
For the higher-order terms
this is not necessary, the convergence in $b$ is fast enough.
We get 
\begin{equation}
 \sigma_{\rm LO} = 2\pi \int_{b_{\rm min}}^{b_{\rm max}} 
 P_{LO}(b) \: b \:  db
 = 2\pi A_{2} \ln \frac{b_{\rm max}}{b_{\rm min}}
\end{equation}
and
\begin{equation}
 \sigma_{\rm NLO}  = 2\pi \int_{b_{\rm min}}^{\infty} 
 P_{NLO}(b) \: b \:  db
 = - \frac{\pi A_{4}}{b_{\rm min}^2} \: .
\end{equation} 
(A somewhat different method was used in eqs.\
(16) and (17) of \cite{tyba95}, where the $\xi$-dependence of the 
first-order amplitude $a_1$ is taken into account. Then there is no need for
the adiabatic cutoff $b_{\rm max}$.) 

We note that the strength parameter $z$ is proportional to $1/v$,
i.e. the leading-order term decreases like $1/E$, the
next-to-leading-order term like 
$1/E^{2}$, where $E$ is the beam energy.
In the halo limit ($\gamma \equiv \gamma_{0},\gamma_{1} \rightarrow 0$) 
the probabilities scale as
$P_{\rm LO} \propto 1/\gamma$ and $P_{\rm NLO} \propto 1/\gamma^3$
($z$ scales as $1/\gamma$) and the excitation probability tends to
infinity. However, the NLO contribution tends to infinity even faster
and higher-order corrections become more important.
Still, for realistic values of $\gamma$ 
%($\gamma_{0}=0.41$ 
%and $\gamma_1=0.25$ for ${}^{11}$Be),
the higher order effects are quite small (see below).
It seems intuitively understandable that
higher-order effects tend to increase with
decreasing $\gamma$. In fact, this increase with
decreasing $\gamma$ is faster than the increase in
lowest order, as our analysis shows.

For the total cross section there is the additional
well known enhancement 
$\ln{\gamma}$ in LO, which
is absent in NLO, due to the fast convergence in $b$.

In \cite{tyba01} and \cite{tyba94} the scaling properties of
the Coulomb dissociation of halo nuclei were investigated
in a related approach.
The strength parameter defined there corresponds
to the one defined now. Using the present notation  we found
in that case that
$P_{\rm LO} \propto 1/\gamma^{3}$
and 
$P_{\rm NLO} \propto 1/\gamma^{5}$.
The difference to the present case arises because the 
bound-continuum and bound-bound state dipole matrix elements show 
different scaling properties with $\gamma_{i}$.

\section{Application to the ${}^{11}$Be Coulomb excitation experiments}
%+ ${}^{208}$Pb 
%experiments at GANIL, MSU, and RIKEN
%and beam energy dependence of $r$ (see Fig.2 \cite{tyba95})}

\begin{table}
\caption{Projectile energy per nucleon $E/A$, velocity $v$, 
  maximum impact parameter $b_{\rm max}$, maximum parameter 
  $z_{\rm max}$ and cross section reduction factor $R(v)$
  for the ${}^{11}$Be experiments.}
\label{tab:1}       % Give a unique label
\begin{tabular}{lccccc}
\hline\noalign{\smallskip}
& $E/A$ & $v$ & $b_{\rm max}$ & $z_{\rm max}$ & $R(v)$ \\
& [MeV] & [$c$] & [fm]          &             &        \\
\noalign{\smallskip}\hline\noalign{\smallskip}
GANIL \cite{xcdcc} & $38.6$ & $0.2861$ & $184.1$ & $0.6549$ & $0.8902$ \\
GANIL \cite{ganil} & $43.0$ & $0.3009$ & $194.6$ & $0.6227$ & $0.9025$ \\
MSU   \cite{fau}   & $60.0$ & $0.3507$ & $230.9$ & $0.5343$ & $0.9321$ \\
RIKEN \cite{riken} & $64.0$ & $0.3611$ & $238.8$ & $0.5189$ & $0.9366$ \\
\noalign{\smallskip}\hline
\end{tabular}
\end{table}

In the  case of  ${}^{11}$Be we have neutron separation
energies of $E_{0} = 504$~keV and
$E_{1}=184$~keV for the ground and excited bound state. 
%in ${}^{11}$Be.
This leads to an excitation energy of
$\hbar \omega=320$~keV for the $1/2^{-}$ state.
With  a radius $R=2.78$~fm one obtains the numerical values 
$\gamma_{0} = 0.4116$ and $\gamma_{1}=0.2487$, respectively, 
for the dimensional
scaling parameters that appear in the excitation amplitude.
Depending on the projectile energy, the
projectile velocities $v$ and the maximum impact parameters
$b_{\max} =\gamma_{\rm beam}v/\omega$ are found for
the conditions of the experiments
at GANIL \cite{xcdcc,ganil}, RIKEN \cite{riken} and MSU \cite{fau}.
They are given in table \ref{tab:1}.
These adiabatic cutoff radii are impressively large.
It means that Coulomb excitation extends really far out and there is 
an amply large region where the interaction 
%everything 
is purely
electromagnetic, and our approach works very 
transparently. Of course,
there is a Coulomb and nuclear interference zone, quite moderate
in extension, close to the minimum impact parameter
$b_{\rm min} = 1.2 (11^{1/3}+208^{1/3})$~fm~$=9.78$~fm
which is not considered in our approach.

The maximum value of the dimensionless strength parameter (\ref{defz})
is given by $z_{\rm max} = 2ZZ_{\rm eff}^{(1)} e^{2} R/
[\hbar v b_{\rm min} (\gamma_{0}+\gamma_{1})]$ 
with the target charge number $Z=82$ and the effective
charge number $Z_{\rm eff}^{(1)} = 4/11$. The numerical values for
$z_{\rm max}$
are given again in table \ref{tab:1}. They are smaller than one
and decrease with the projectile energy. The range where the reduction
factor is actually
needed is well approximated by the NLO
correction, which corresponds to the approximation of $r(z)$ in figure
\ref{fig:1} by an inverted parabola. 
In the present approach we are more certain
about the nuclear structure input that determines the opening
parameter of this parabola. 

Similar as in \cite{tyba95}
we define a reduction factor $R(v) = \sigma^{(\infty)}/\sigma_{LO}$ 
of the total excitation cross section 
$\sigma^{(\infty)}$ caused by higher-order effects.
This factor $R(v)$ can be used to take higher order effects
into account in an analysis of the experimental data. 

\begin{figure}
%\resizebox{0.75\textwidth}{!}{%
\resizebox{0.48\textwidth}{!}{%
\includegraphics{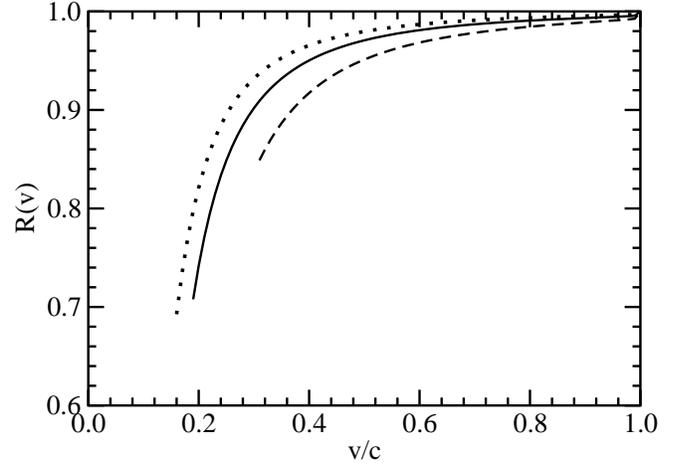}
}
\caption{Reduction factor $R(v)=1+\sigma_{NLO}/\sigma_{LO}$
as a function of the velocity $v$ for values of $0$ (dashed line), 
$0.604$ (solid line) and $1$ (dotted line) of the ratio 
$\gamma_{1}/\gamma_{0} = \sqrt{E_{1}/E_{0}}$.
}
\label{fig:2}       % Give a unique label
\end{figure}

Since the
parameter $z$ is always smaller than one in the experiments, 
the full cross section
in sudden approximation $\sigma^{(\infty)} = \sigma_{LO} + 
\sigma_{NLO} + \dots $
is well approximated by the NLO correction and the reduction factor
becomes
\begin{equation}
 R(v) \approx 1 + \frac{\sigma_{NLO}}{\sigma_{LO}}
  = 1 - \frac{A_{4}}{2A_{2}b_{\rm min}^{2}\ln \frac{b_{\rm
        max}}{b_{\rm min}}} \: .
\end{equation}
The numerical values as given in table \ref{tab:1} show that
the reduction of the first-order cross section is in the order
of 6 to 11~\% for the various experiments. In figure \ref{fig:2}
the dependence of the reduction on the projectile velocity $v$
is depicted for the actual value of the ratio $\gamma_{1}/\gamma_{0}$
and the two limiting cases as in figure \ref{fig:1}. The curves are
shown only for velocities where the parameter $z$ does not exceed
the value one, the radius of convergence for the expansion of the
amplitude (\ref{asuddenseries}). Note that an increase of $\gamma_{1}$
from zero (dashed line) to one (dotted line) leads to a reduction
of the parameter $z$ and correspondingly to a reduction of the
higher-order corrections in the cross section for the same velocity $v$.

\section{Conclusion}

We have presented a realistic model
for higher-order effects in the Coulomb excitation
of neutron halo nuclei.
With quite simple methods a reliable value for the reduction
of the cross section could be given.
In addition, this should be very useful as a benchmark, 
for valuable and necessary tests of more sophisticated approaches
like CDCC or time-dependent calculations. 
These kinds of calculations require a lot of 
computation, yet they should reproduce the present 
results in certain well defined limits. 
Using an effective-range point of view
we updated our previous work \cite{tyba95}.
Other angular momentum combinations can be treated in a similar way.
However, for higher angular momenta the halo nature is less pronounced
and the importance of the core size parameter R increases.
Higher-order terms in $\gamma$ will become more important and the 
method less useful.
It is quite remarkable that our results for 
${}^{11}$Be Coulomb excitation depend only on 
the binding energies (they determine the parameters $q_{0}$ and $q_{1}$)
and a parameter $R$ which characterizes the core size.
This stresses the fact that halo nuclei depend only on a few
low-energy constants, and not on details of the shape of the 
potential. This is in accord with low-energy scattering,
which is also characterized by a few low-energy parameters
in the effective-range expansion. 

In view of the discovery of the neutron-rich
isotopes ${}^{40}$Mg and ${}^{42}$Al \cite{msu,Hee07} one may
expect that more neutron-rich nuclei with low-lying single
particle states exist to which the present approach
can be applied. It is very likely that the new radioactive beam 
facilities around the world like FAIR, SPIRAL2 and RIBF
will provide interesting new examples
in the higher mass region.

%%
%% For two-column wide figures use
%\begin{figure*}
%% Use the relevant command for your figure-insertion program
%% to insert the figure file. See example above.
%% If not, use
%\vspace*{5cm}       % Give the correct figure height in cm
%\end{figure*}
%\begin{table}
%\begin{tabular}{lll}
%\hline\noalign{\smallskip}
%\end{tabular}
%% Or use
%\vspace*{5cm}  % with the correct table height
%\end{table}

\section*{Acknowledgment}

This research was supported by the DFG cluster of excellence
``Origin and Structure of the Universe''.

\section*{Appendix}
For the analytical calculation of the excitation amplitude in 
sudden approximation the argument of the exponential function
in (\ref{asudden}) can be written as $\vec{q}_{\rm Coul}\cdot\vec{r} =
q_{\rm Coul}r\cos(\theta)$
in spherical coordinates because the vector $\vec{q}_{\rm Coul}$
points in $z$-direction. Using
\begin{equation}
\left[ \mathcal{Y}_{j_{f}m_{f}}^{l_{f}s}(\hat{r}) \right]^{\dagger}
 \mathcal{Y}_{j_{i}m_{i}}^{l_{i}s}(\hat{r})
 = (-1)^{j_{i}+m_{i}} \delta_{m_{i}m_{f}} \frac{\cos(\theta)}{4\pi}
\end{equation}
for $j_{i} = j_{f} = |m_{i}| = |m_{f}| = \frac{1}{2}$, 
$l_{i} = 0$, $l_{f} = 1$ and $s=1/2$, one finds
\begin{eqnarray}
 \lefteqn{a_{fi}} \\ \nonumber 
 & = & (-1)^{j_{i}+m_{i}} \frac{C_{0} C_{1}}{4\pi} 
 \int_{0}^{\infty} dr \int_{0}^{2\pi} d\phi \int_{-1}^{1}
 d\cos(\theta) \: \cos(\theta)
 \\ \nonumber & & \times
  \exp[-(q_{1}+q_{0})r]  \left( 1 + \frac{1}{q_{1}r} \right)
 \exp[-iq_{\rm Coul}r\cos(\theta)] 
 \\ \nonumber & = & 
 (-1)^{j_{i}+m_{i}} \frac{C_{0} C_{1}}{2i} \left( \frac{1}{q_{\rm Coul}} I_{1} 
 + \frac{1}{q_{1}} I_{2}\right)
\end{eqnarray}
with two integrals that can be calculated separately.
%(In the case with $m_{i} = m_{f} = -1/2$ the product of the spinor
%spherical harmonics and the amplitude change the sign.)
The first integral is given by
\begin{eqnarray}
  I_{1}(p,q) & = & iq \int_{0}^{\infty} dr \int_{-1}^{1} dz \: z  
  \exp[-(p+iqz)r]
 \\ \nonumber & = & iq \int_{-1}^{1} dz \: 
 \frac{z}{p+iqz}
 \\ \nonumber & = &
   2 - \frac{p}{iq} \ln \frac{p+iq}{p-iq}
\end{eqnarray}
with $p = q_{0}+q_{1}$ and $q=q_{\rm Coul}$. With the relation
\begin{equation}
 \ln \frac{1+iz}{1-iz} = 2i \arctan (z)
\end{equation}
one obtains
\begin{equation}
  I_{1}(p,q)  =  2
  \left[ 1 - \frac{p}{q} \arctan\left( \frac{q}{p} \right)\right] \: .
\end{equation}
The second integral is given by
%\begin{equation}
%  I_{2}(p,q) = i \int_{0}^{\infty} dr \int_{-1}^{1} dz \: z
%  \frac{\exp[-(p+iqz)r]}{r} \: .
%\end{equation}
\begin{eqnarray}
  I_{2}(p,q) & = & i \int_{0}^{\infty} dr \int_{-1}^{1} dz \: z
  \frac{\exp[-(p+iqz)r]}{r}  
 \\ \nonumber & = & 
 - \frac{d}{dq} \int_{0}^{\infty} dr \int_{-1}^{1} dz \: 
  \frac{\exp[-(p+iqz)r]}{r^{2}} 
 \\ \nonumber & = & 
 2 \int_{0}^{\infty} dr \: \frac{\exp(-pr)}{r} j_{1}(qr)
\end{eqnarray}
with a spherical Bessel function $j_{1}(x) = - \frac{d}{dx} j_{0}(x)$ so that
\begin{equation}
 \lim_{q \to 0} I_{2}(p,q) = 0 \: .
\end{equation}
We notice
\begin{equation}
 \frac{d}{dp} I_{2}(p,q) = - \frac{1}{q} I_{1}(p,q)
\end{equation}
and find with the help of a partial integration
\begin{eqnarray}
 I_{2}(p,q) & = & C - \frac{1}{q} \int_{0}^{p} dp^{\prime} \: 
 I_{1}(p^{\prime},q)
 \\ \nonumber & =  & 
 C - 2 \frac{p}{q} + \frac{p^{2}}{q^{2}}
\arctan\left( \frac{q}{p} \right) 
 \\ \nonumber &  & 
 +  \frac{p}{q} - \arctan\left( \frac{p}{q}\right) 
\end{eqnarray}
where a constant $C$ appears.
Since \begin{equation}
 \arctan (x) + \arctan\left(\frac{1}{x}  \right) = \frac{\pi}{2}
\end{equation}
for $x \ge 0$
we can write
\begin{equation}
 I_{2}(p,q)  = 
 C -  \frac{p}{q} + \left( 1 + \frac{p^{2}}{q^{2}} \right)
\arctan\left( \frac{q}{p} \right) 
 - \frac{\pi}{2} \: .
\end{equation}
With
\begin{equation}
 \lim_{q \to 0} I_{2}(p,q) = C - \frac{\pi}{2}  = 0 
\end{equation}
the constant can be determined to be $C= \pi/2$
and the final result is
\begin{equation}
 I_{2}(p,q)  = 
 \left( 1 + \frac{p^{2}}{q^{2}} \right)
\arctan\left( \frac{q}{p} \right)  -  \frac{p}{q} \: .
\end{equation}

% BibTeX users please use
% \bibliographystyle{}
% \bibliography{}

\end{document}